\documentclass[10pt,a4paper,onecolumn]{article}
\usepackage{marginnote}
\usepackage{graphicx}
\usepackage{xcolor}
\usepackage{authblk,etoolbox}
\usepackage{titlesec}
\usepackage{calc}
\usepackage{tikz}
\usepackage{hyperref}
\hypersetup{colorlinks,breaklinks,
            urlcolor=[rgb]{0.0, 0.5, 1.0},
            linkcolor=[rgb]{0.0, 0.5, 1.0}}
\usepackage{caption}
\usepackage{tcolorbox}
\usepackage{amssymb,amsmath}
\usepackage{ifxetex,ifluatex}
\usepackage{seqsplit}
\usepackage{fixltx2e} 
\usepackage[
  backend=biber,
]{biblatex}

\usepackage[top=3.5cm, bottom=3cm, right=1.5cm, left=1.0cm,
            headheight=2.2cm, reversemp, includemp, marginparwidth=4.5cm]{geometry}

\titleformat{\section}
  {\normalfont\sffamily\Large\bfseries}
  {}{0pt}{}
\titleformat{\subsection}
  {\normalfont\sffamily\large\bfseries}
  {}{0pt}{}
\titleformat{\subsubsection}
  {\normalfont\sffamily\bfseries}
  {}{0pt}{}
\titleformat*{\paragraph}
  {\sffamily\normalsize}

\usepackage{fancyhdr}
\makeatletter
\let\ps@plain\ps@fancy
\fancyheadoffset[L]{4.5cm}
\fancyfootoffset[L]{4.5cm}

\definecolor{linky}{rgb}{0.0, 0.5, 1.0}

\newtcolorbox{repobox}
   {colback=red, colframe=red!75!black,
     boxrule=0.5pt, arc=2pt, left=6pt, right=6pt, top=3pt, bottom=3pt}

\newcommand{\ExternalLink}{%
   \tikz[x=1.2ex, y=1.2ex, baseline=-0.05ex]{%
       \begin{scope}[x=1ex, y=1ex]
           \clip (-0.1,-0.1)
               --++ (-0, 1.2)
               --++ (0.6, 0)
               --++ (0, -0.6)
               --++ (0.6, 0)
               --++ (0, -1);
           \path[draw,
               line width = 0.5,
               rounded corners=0.5]
               (0,0) rectangle (1,1);
       \end{scope}
       \path[draw, line width = 0.5] (0.5, 0.5)
           -- (1, 1);
       \path[draw, line width = 0.5] (0.6, 1)
           -- (1, 1) -- (1, 0.6);
       }
   }

\patchcmd{\@maketitle}{center}{flushleft}{}{}
\patchcmd{\@maketitle}{center}{flushleft}{}{}
\patchcmd{\@maketitle}{\LARGE}{\LARGE\sffamily}{}{}
\def\maketitle{{%
  
  \AB@maketitle}}
\makeatletter
\renewcommand\AB@affilsepx{ \protect\Affilfont}
\renewcommand\AB@affilnote[1]{{\bfseries #1}\hspace{3pt}}
\makeatother

\renewcommand\Affilfont{\sffamily\small\mdseries}
\ifnum 0\ifxetex 1\fi\ifluatex 1\fi=0 
  \usepackage[T1]{fontenc}
  \usepackage[utf8]{inputenc}

\else 
  \ifxetex
    \usepackage{mathspec}
  \else
    \usepackage{fontspec}
  \fi
  \defaultfontfeatures{Ligatures=TeX,Scale=MatchLowercase}

\fi
\IfFileExists{upquote.sty}{\usepackage{upquote}}{}
\IfFileExists{microtype.sty}{%
\usepackage{microtype}
\UseMicrotypeSet[protrusion]{basicmath} 
}{}

\usepackage{hyperref}
\hypersetup{unicode=true,
            pdftitle={powerbox: A Python package for creating structured fields with isotropic power spectra},
            pdfborder={0 0 0},
            breaklinks=true}
\usepackage{graphicx,grffile}
\makeatletter
\def\maxwidth{\ifdim\Gin@nat@width>\linewidth\linewidth\else\Gin@nat@width\fi}
\def\maxheight{\ifdim\Gin@nat@height>\textheight\textheight\else\Gin@nat@height\fi}
\makeatother
\setkeys{Gin}{width=\maxwidth,height=\maxheight,keepaspectratio}
\IfFileExists{parskip.sty}{%
\usepackage{parskip}
}{
\setlength{\parindent}{0pt}
\setlength{\parskip}{6pt plus 2pt minus 1pt}
}
\ifx\paragraph\undefined\else
\let\oldparagraph\paragraph
\renewcommand{\paragraph}[1]{\oldparagraph{#1}\mbox{}}
\fi
\ifx\subparagraph\undefined\else
\let\oldsubparagraph\subparagraph
\renewcommand{\subparagraph}[1]{\oldsubparagraph{#1}\mbox{}}
\fi

\title{powerbox: A Python package for creating structured fields with isotropic
power spectra}

        \author[1, 2]{Steven G. Murray}
    
      \affil[1]{International Centre for Radio Astronomy Research (ICRAR), Curtin
University, Bentley, WA 6102, Australia}
      \affil[2]{ARC Centre of Excellence for All-Sky Astrophysics in 3 Dimensions (ASTRO
3D)}
  \date{\vspace{-5ex}}

\begin{document}
\maketitle

\marginpar{
  \sffamily\small

  {\bfseries DOI:} \href{https://doi.org/10.21105/joss.00850}{\color{linky}{10.21105/joss.00850}}

  \vspace{2mm}

  {\bfseries Software}
  \begin{itemize}
    \setlength\itemsep{0em}
    \item \href{https://github.com/openjournals/joss-reviews/issues/28}{\color{linky}{Review}} \ExternalLink
    \item \href{https://github.com/steven-murray/powerbox}{\color{linky}{Repository}} \ExternalLink
    \item \href{http://dx.doi.org/10.21105/zenodo.1400822}{\color{linky}{Archive}} \ExternalLink
  \end{itemize}

  \vspace{2mm}

  {\bfseries Submitted:} 23 July 2018\\
  {\bfseries Published:} 21 August 2018

  \vspace{2mm}
  {\bfseries License}\\
  Authors of papers retain copyright and release the work under a Creative Commons Attribution 4.0 International License (\href{http://creativecommons.org/licenses/by/4.0/}{\color{linky}{CC-BY}}).
}

\hypertarget{summary}{%
\section{Summary}\label{summary}}

The power spectrum is a cornerstone of both signal analysis and spatial
statistics, encoding the variance of a signal or field on different
scales. Its common usage is in no small part attributable to the fact
that it is a \emph{full} description of a purely Gaussian process -- for
such statistical processes, no information is contained in higher-order
statistics. The prevalence of such processes (or close approximations to
them) in physical systems serves to justify the popularity of the power
spectrum as a key descriptive statistic in various physical sciences,
eg. cosmology (Peacock 1998) and fluid mechanics (Monin and Yaglom
2007). It furthermore readily avails itself to efficient numerical
evaluation, being the absolute square of the Fourier Transform.

Another feature of many approximate physical systems, especially those
already mentioned, is that they are both homogeneous and isotropic (at
least in some local sample). In this case, the \emph{n}-dimensional
power spectrum may be losslessly compressed into a single dimension,
which is radial in Fourier-space. Such processes approximately describe
for example the over-density field of the early Universe and locally
isotropic turbulent flows. Thus it is of great use to have a numerical
code which simplifies the dual operations of; (i) producing random
homogeneous/isotropic fields (of arbitrary dimensionality) consistent
with a given 1D radial power spectrum, and (ii) determination of the 1D
radial power spectrum of random fields (or a sample of tracers of that
field). \texttt{powerbox} exists to perform these duals tasks with both
simplicity and efficiency.

Performing the first of these tasks is especially non-trivial. While the
power spectrum can be evaluated on any field (though it may not fully
describe the given field), the precise machinery for \emph{creating} a
field from a given power spectrum depends on the probability density
function (PDF) of the process itself. The machinery for creating a
Gaussian field is well-known. However, other PDF's -- especially those
that are positively bounded -- are extremely useful for describing such
physical entities as density fields. In these cases, the
\emph{log-normal} PDF has become a standard approximation (Coles and
Jones 1991), and \texttt{powerbox} makes a point of supporting the
machinery for generating log-normal fields (Beutler et al. 2011) for
this purpose. Indeed, \texttt{powerbox} is \emph{primarily} geared
towards supporting cosmological applications, such as measuring and and
producing samples of galaxy positions in a log-normal density field
(while account for standard effects such as shot-noise and standard
normalisation conventions). It is nevertheless flexible enough to
support research in any field (with its own particular conventions) that
is based on the homogeneous and isotropic power spectrum.

\texttt{Powerbox} is a pure-Python package devoted to the simple and
efficient solution of the previous considerations. As the most popular
language for astronomy, Python is the natural language of choice for
\texttt{powerbox}, with its focus on cosmological applications, and it
also provides for great ease-of-use and extensibility. As an example of
the former, all functions/classes within \texttt{powerbox} are able to
work in arbitrary numbers of dimensions (memory permitting), simply by
setting a single parameter \emph{n}. As an example of the latter, the
class-based structure of the field-generator may be used to extend the
generation to fields with PDF's other than either Gaussian or log-normal
(indeed, the log-normal class is itself sub-classed from the Gaussian
one). \texttt{powerbox} does not sacrifice efficiency for its high-level
interface. By default, the underlying FFT's are performed by
\texttt{numpy}, which uses underlying fast C code. In addition, if the
\texttt{pyFFTW} package is installed, \texttt{powerbox} will seamlessly
switch to using its optimized C code for up to double the efficiency. It
is also written with an eye for conserving memory, which is important
for the often very large fields that may be required.

\texttt{Powerbox} was written due to research-demand, and as such it is
highly likely to be suited to the requirements of research of a similar
nature. Furthermore, as previously stated, every effort has been made to
sufficiently generalize its scope to be of use in related fields of
research. It has already been instrumental in several publications
(Murray, Trott, and Jordan 2017; Wolz et al. 2018), and we hope it will
be a useful tool for approximate theoretical simulations by many others.

\hypertarget{acknowledgements}{%
\section{Acknowledgements}\label{acknowledgements}}

The author acknowledges helpful discussions and contributions from
Cathryn Trott, Chris Jordan and Laura Wolz during the initial
development of this project. Parts of this research were supported by
the Australian Research Council Centre of Excellence for All Sky
Astrophysics in 3 Dimensions (ASTRO 3D), through project number
CE170100013

\hypertarget{references}{%
\section*{References}\label{references}}
\addcontentsline{toc}{section}{References}

\hypertarget{refs}{}
\leavevmode\hypertarget{ref-Beutler2011}{}%
Beutler, Florian, Chris Blake, M. Colless, D. H. Jones, L.
Staveley-Smith, Lachlan A. Campbell, Q. Parker, W. Saunders, and F.
Watson. 2011. ``The 6dF Galaxy Survey: Baryon Acoustic Oscillations and
the Local Hubble Constant.'' \emph{Mon. Not. R. Astron. Soc.} 416 (4):
3017--32. \url{https://doi.org/10.1111/j.1365-2966.2011.19250.x}.

\leavevmode\hypertarget{ref-Coles1991}{}%
Coles, Peter, and Bernard Jones. 1991. ``A Lognormal Model for the
Cosmological Mass Distribution.'' \emph{Mon. Not. R. Astron. Soc.} 248:
1--13. \url{https://doi.org/10.1093/mnras/248.1.1}.

\leavevmode\hypertarget{ref-Monin2007}{}%
Monin, A. S., and A. M. Yaglom. 2007. \emph{Statistical Fluid
Mechanics}. Vol. 1. Dover Books on Physics. New York: Dover
Publications.

\leavevmode\hypertarget{ref-Murray2017}{}%
Murray, S. G., C. M. Trott, and C. H. Jordan. 2017. ``An Improved
Statistical Point-Source Foreground Model for the Epoch of
Reionization.'' \emph{ApJ} 845 (1): 7.
\url{https://doi.org/10.3847/1538-4357/aa7d0a}.

\leavevmode\hypertarget{ref-Peacock1999}{}%
Peacock, J. A. 1998. ``Cosmological Density Fields.'' In
\emph{Cosmological Physics}, 495--552. Cambridge University Press.
\url{https://doi.org/10.1017/CBO9780511804533.017}.

\leavevmode\hypertarget{ref-Wolz2018}{}%
Wolz, L., S. G. Murray, C. Blake, and J. S. Wyithe. 2018. ``Intensity
Mapping Cross-Correlations II: HI Halo Models Including Shot Noise.''
\emph{ArXiv180302477 Astro-Ph}, March.
\url{http://arxiv.org/abs/1803.02477}.

\end{document}